\documentclass[preprint,12pt,authoryear]{elsarticle}

\usepackage{color}
\usepackage{amsmath,amsfonts}
\usepackage{color}
\usepackage{graphicx}

\usepackage{geometry}
\newcommand{\be}{\begin{equation}}
\newcommand{\bea}{\begin{eqnarray}}
\newcommand{\ee}{\end{equation}}
\newcommand{\eea}{\end{eqnarray}}

\newcommand{\calA}{\mathcal{A}}
\newcommand{\calC}{\mathcal{C}}

\newcommand{\calS}{\mathcal{S}}
\newcommand{\calL}{\mathcal{L}}

\newcommand{\dr}{\,\text{d}r}
\newcommand{\dtheta}{\,\text{d}\theta}
\newcommand{\drdtheta}{\,\text{d}r\mskip1mu\text{d}\theta}
\newcommand{\drhodtheta}{\,\text{d}\rho\mskip1.75mu\text{d}\theta}

\newcommand{\lint}{\int_0^{2\pi}}
\newcommand{\aint}{\int_0^{2\pi}\mskip-8mu\int_0^R}
\newcommand{\aintdim}{\int_0^{2\pi}\mskip-8mu\int_0^1}

\newcommand{\trans}{\mskip-2mu\scriptscriptstyle\top\mskip-2mu}

\newcommand{\half}{{\textstyle{\frac{1}{2}}}}

\newcommand{\bfa}{\boldsymbol{a}}
\newcommand{\bfb}{\boldsymbol{b}}

\newcommand{\bfd}{\boldsymbol{d}}
\newcommand{\bfe}{\boldsymbol{e}}
\newcommand{\bff}{\boldsymbol{f}}
\newcommand{\bfm}{\boldsymbol{m}}
\newcommand{\bfn}{\boldsymbol{n}}
\newcommand{\bfp}{\boldsymbol{p}}

\newcommand{\bft}{\boldsymbol{t}}
\newcommand{\bfu}{\boldsymbol{u}}
\newcommand{\bfv}{\boldsymbol{v}}
\newcommand{\bfx}{\boldsymbol{x}}

\newcommand{\bflambda}{\boldsymbol{\lambda}}
\newcommand{\bfsigma}{\boldsymbol{\sigma}}

\newcommand{\bfxi}{\boldsymbol{\xi}}

\newcommand{\bfP}{\boldsymbol{P}}
\newcommand{\bfQ}{\boldsymbol{Q}}

\newcommand{\bfT}{\boldsymbol{T}}

\newcommand{\bfM}{\boldsymbol{M}}

\newcommand{\bfeperp}{\boldsymbol{e}^{\scriptscriptstyle\perp}}

\newfont{\tenbfsl}{cmbxti12 at 13 pt}
\newcommand{\idem}{\mbox{\tenbfsl 1\/}}

\newcommand{\bbR}{\mathbb{R}}

\newcommand{\eql}{\scriptstyle\mskip1.75mu=\mskip1.75mu}

\newcommand{\ctimes}{\mskip-1.25mu\times\mskip-1.25mu}

\usepackage{amssymb}

\journal{Journal of the Mechanics and Physics of Solids}

\begin{document}

\begin{frontmatter}



\title{Stability and bifurcation of a soap film spanning an elastic loop}


\author[label1]{Yi-chao Chen}
\author[label2]{Eliot Fried}

\address[label1]{Department of Mechanical Engineering, University of Houston, Houston, TX 77204-4006, USA}
\address[label2]{Department of Mechanical Engineering, McGill University, Montr\'eal, Qu\'ebec H3A 2K6, Canada}

\begin{abstract}
The Euler--Plateau problem, proposed by \cite{gm}, concerns a soap film spanning a flexible loop. The shapes of the film and the loop are determined by the interactions between the two components. In the present work, the Euler--Plateau problem is reformulated to yield a boundary-value problem for a vector field that parameterizes both the spanning surface and the bounding loop. Using the first and second variations of the relevant free-energy functional, detailed bifurcation and stability analyses are performed. For spanning surface with energy density $\sigma$ and a bounding loop with length $2\pi R$ and bending rigidity $a$, the first bifurcation, during which the spanning surface remains flat but the bounding loop becomes noncircular, occurs at $\sigma R^3/a=3$, confirming a result obtained previously via an energy comparison. Other bifurcation solution branches, including those emanating from the flat circular solution branch to nonplanar solution branches, are also shown to be unstable.
\end{abstract}


\end{frontmatter}

\section{Introduction}

What is known as the ``thread problem" involves determining one or more surfaces with (pointwise) zero mean-curvature that spans a closed loop consisting of a rigid segment of finite length joined at its ends by a flexible segment of prescribed length. An essential part of finding a solution to the thread problem is to determine the shape of the flexible segment of the bounding loop. This problem has attracted considerable attention from mathematicians. See \citet[\S5.4]{dht} for a synopsis of the  salient developments.

The thread problem embraces two complementary specializations. If the flexible segment of the bounding loop is empty, so that the boundary consists of a closed rigid loop of prescribed shape, the thread problem reduces to the classical Plateau problem, which occupied the efforts of some of the most prominent mathematicians of 19th and 20th centuries. See \citet[\S4.15]{dhs} for a comprehensive overview of the formative contributions.

Scarcely any attention has been given to the alternative specialization, in which the rigid segment of the bounding loop is empty, so that the boundary is a closed flexible loop of fixed length but indeterminate shape. Indeed, the literature appears to contain only two papers in which this problem is considered, one published by \cite{by} in 2001 and another published much more recently by \cite{gm} in 2012. Two earlier papers, published by \cite{b1,b2} in 1997 addressed closely related problems in arbitrary dimensions and codimensions.

Following the conventional formulation of the Plateau problem, \cite{by} and  Giomi \& Mahadevan~\cite{gm} adopt variational approaches in which the spanning surface is endowed with a constant free-energy density and the bounding loop is endowed with a free-energy density quadratic in some measure of curvature. In each case, the resulting model involves a competition between areal and lineal contributions to the net free-energy. The particular features of the lineal free-energy densities that \cite{by} and \cite{gm} are, however, somewhat different.

\cite{by} measure the free-energy of the spanned bounding loop relative to a reference state corresponding to the shape of the unspanned loop. In the Bernoulli--Euler theory of thin beams, this would be achieved by introducing a free-energy density proportional to the square of the difference between the curvature of the spanned bounding loop and the intrinsic curvature of the unspanned loop. At variance with this, \cite{by} use an expression that is proportional to the square of the magnitude of the difference between the curvature vectors of the spanned and unspanned loops. In terms of the corresponding scalar measures of curvature, their expression can be written as the sum of a  Bernoulli--Euler term accounting for intrinsic curvature and an unconventional term dependent on the angle between the two curvature vectors. For their net free-energy, \cite{by} establish the existence of minimizers involving bounding curve with no points of self-contact. Underlying this result is a hypothesis limiting the shape of the unspanned bounding loop and sufficient to rule out minimizers with points of self-contact.

In contrast, \cite{gm} adopt Bernoulli--Euler theory and model the bounding loop as an inextensible, twist-free elastica with constant flexural rigidity $a>0$ and no intrinsic curvature. If the spanning surface has constant surface tension $\sigma>0$ and the bounding loop is of length $L=2\pi R$, what results is a model involving a single dimensionless parameter
\be
\gamma=\frac{16\pi^3R^3\sigma}{a}
\label{gammaintro}
\ee
that measures the importance of areal free-energy relative to lineal free-energy. To explore the competition between these contributions to the net free-energy, \cite{gm} devised an innovative modification of Plateau's \citeyearpar{p} experiments with soap films. This involves dipping and extracting circular loops made of different lengths of fishing line into soapy water, while holding fixed the diameter and flexural rigidity of the fishing line and the surface free-energy of the soapy water. As the length $L$ of the bounding contour, and consequently $\gamma$, increases, various remarkable shape transitions are observed. Sufficiently short loops are spanned by flat disks. Loops of intermediate lengths are spanned by nonplanar, saddle-like surfaces. Sufficiently long loops adopt twisted shapes exhibiting self-contact.

Relying on an energy comparison argument, \cite{gm} find that flat disks with circular boundaries become unstable at $\gamma=48\pi^3$, giving way to flat spanning surfaces bounded by noncircular loops. To investigate transitions to nonplanar configurations, they use a combination of numerical and asymptotic methods. On these grounds, a supercritical pitchfork bifurcation from plane to out-of-plane configurations is predicted to occur at $\gamma=96\pi^3$.

On computing and setting to zero the first variation of their net free-energy, \cite{gm} find that, consistent with what occurs in the Plateau and thread problems, the mean curvature $H$ of spanning surface must vanish pointwise:
\be
H=0.
\label{Heq0intro}
\ee
Additionally, they arrive at a pair of conditions involving the curvature $\kappa$ and torsion $\tau$ of the bounding curve and the angle $\vartheta$ between unit normal of the Frenet frame of the bounding loop and the restriction to the bounding loop of the unit normal to the spanning surface. On using $s$ to measure arclength along the bounding loop, these conditions are
\be
\kappa_{ss}+\frac{1}{2}\kappa^3-\bigg(\tau^2+\frac{\beta}{a}\bigg)\kappa-\frac{\sigma}{a}\sin\vartheta=0
\label{b2intro}
\ee
and
\be
2\kappa_s\tau+\kappa\mskip1mu\tau_s+\frac{\sigma}{a}\cos\vartheta=0,
\label{b1intro}
\ee
where a subscript indicates differentiation with respect to $s$ and $\beta$ is a constant Lagrange multiplier associated with the constraint that the bounding loop be inextensible.  In \eqref{b2intro} and \eqref{b1intro}, the terms $-\sigma\mskip1.5mu\sin\vartheta$ and $\sigma\mskip1.5mu\cos\vartheta$ represent the forces exerted by the spanning surface on the bounding loop in the normal and binormal directions of the Frenet frame. Consistent with this observation, formally setting $\sigma$ to zero reduces \eqref{b2intro} and \eqref{b1intro} to the equilibrium equations for an inextensible, twist-free elastica; see, for example, equations (1) and (2) in Singer's \citeyearpar{singer} lectures on elastic curves and rods.

Despite their interesting and appealing geometrical attributes, the first-variation conditions \eqref{Heq0intro}--\eqref{b1intro} do not provide a well-posed boundary-value problem for finding the shapes of the spanning surface and the bounding loop. To clarify this assertion, suppose that the contact angle $\vartheta$ is known and that the system \eqref{b2intro}--\eqref{b1intro} has been solved to determine the curvature $\kappa$ and torsion $\tau$ as smooth functions, with common period, of the arclength $s$. By the fundamental theorem of curves \citep[pp.~19--20, pp.~309--311]{dc}, $\kappa$ and $\tau$ can then be used to construct a space curve, unique up to a rigid transformation. Their periodicity does not, however, ensure that the curve they generate is closed. Indeed, the problem of obtaining explicit conditions conditions on $\kappa$ and $\tau$ necessary to ensure that they generate a closed curve, the statement of which is commonly attributed to \cite{e} and \cite{f}, remains open. Regarding the full system \eqref{Heq0intro}--\eqref{b1intro}, it also seems reasonable to expect further complications due to the presence of $\vartheta$, which couples the problems of determining the shapes of the spanning surface and the bounding loop. Perhaps due to these difficulties, \cite{gm} opt for a numerically-based energy minimization scheme instead of solving \eqref{Heq0intro}--\eqref{b1intro}.

The aim of the work presented here is to derive the equations of equilibrium and stability conditions through a rigorous analysis which, among other things, circumvents potentially insurmountable challenges aspects of the system \eqref{Heq0intro}--\eqref{b1intro}. The paper is organized as follows. The formulation appears in Section~\ref{formulation}. Aside from adopting the modeling assumptions of \cite{gm}, the spanning surface and bounding loop are represented in terms of a suitably smooth and injective map $\bfx$ defined on the closed disk of radius $R$ and express the net free-energy $F$ as a functional of this mapping. Scaling issues are discussed in Section~\ref{scaling}. The first and second variations $\delta F$ and $\delta^2\mskip-2mu F$ of the net free-energy functional are computed in Section~\ref{energystablity}. In particular, on letting $\bfn$ denote a unit normal to the spanning surface, introducing polar coordinates $r$ and $\theta$, and using subscripts to denote partial derivatives, the first variation condition $\delta F=0$ yields a second-order partial-differential equation
\be
\bfx_\theta\ctimes\bfn_r+\bfn_\theta\ctimes\bfx_r={\bf0},
\qquad
\bfn=\frac{\bfx_r\ctimes\bfx_\theta}{|\bfx_r\ctimes\bfx_\theta|},
\label{16intro}
\ee
that holds pointwise for all $r<R$ and $0\le\theta\le2\pi$, and a fourth-order ordinary-differential equation
\be
[\nu\mskip1mu\bfx_\theta\ctimes\bfn+
(\bfx_{\theta\theta\theta}
-\lambda\mskip1.75mu\bfx_\theta)_\theta]_{r\eql R}={\bf0},
\qquad
\nu=\frac{R^3\sigma}{a},
\label{17intro}
\ee
that holds for $0\le\theta\le2\pi$. In \eqref{17intro}, $\lambda$ is a Lagrange multiplier defined on $0\le\theta\le2\pi$ and needed to ensure the inextensibility of the bounding loop. The geometric content of conditions \eqref{16intro}--\eqref{17intro} is explored in Section~\ref{geometriccontent}. Importantly, \eqref{16intro} embodies the requirement that the vector mean-curvature of the spanning surface vanishes pointwise and, thus, is equivalent to \eqref{Heq0intro}. Additionally, the components of \eqref{17intro} in the normal and binormal directions of the Frenet frame of the bounding loop are equivalent to \eqref{b2intro} and \eqref{b1intro}, respectively. The remaining component, in the direction tangent to the bounding loop, determines the Lagrange multiplier as a function of the curvature $\kappa$. Connections between the equilibrium conditions and relevant force and moment balances for the system are established in Section~\ref{kineticcontent}. Distinguishing characteristics of the boundary-value problem comprised by the equilibrium conditions \eqref{16intro}--\eqref{17intro} are discussed in Section~\ref{challenges}. To facilitate our analysis of stability and bifurcation, a nondimensional version of our formulation is presented in Section~\ref{nondimensionalization}. A stability analysis of the flat circular solution is given in Section~\ref{stability}. It is shown that whether the second variation condition holds depends on the sign of an associated eigenvalue. This eigenvalue problem is then solved, yielding pointwise stability conditions that restrict explicitly the value of $\nu$ defined in \eqref{17intro}. Section~\ref{bifurcation} is devoted to a bifurcation analysis for the flat circular solution branch. By solving the linearized equations of equilibrium, a sequence of bifurcation points are identified. However, all the bifurcating solution branches are found to be unstable, except for that corresponding to the first bifurcation mode, namely the bifurcation from the flat circular solution to a flat solution bounded by a noncircular loop.

\section{Formulation}
\label{formulation}

Consider a soap film bounded by a flexible, but inextensible loop with prescribed length $L=2\pi R$. Following \cite{gm}, the soap film is modeled as a surface $\calS$ with constant free-energy density $\sigma>0$ and the bounding loop as an inextensible, twist-free elastic curve $\calC=\partial\calS$ with constant flexural rigidity $a>0$ and no intrinsic curvature. Then, using $\kappa$ to denote the curvature of $\calC$, the net free-energy of the system consisting of the soap film and bounding loop consists of a sum
\be
F=\int_{\calS}\sigma+\int_{\calC}\half a\kappa^2
\label{energy0}
\ee
of areal and lineal contributions.

Let $r$ and $\theta$ denote polar coordinates, with respective $\theta$-dependent orthonormal base vectors $\bfe$ and $\bfeperp$, and suppose that $\calS$ admits a parametrization
\be
\calS=\{\bfx\in\bbR^3:\bfx=\bfx(r,\theta),0\le r\le R,0\le\theta\le2\pi\},
\label{parametrization}
\ee
with $\bfx$ being a three-times continuously differentiable, injective mapping satisfying closure conditions
\be
\bfx(r,0)=\bfx(r,2\pi)
\qquad\text{and}\qquad
\bfx_\theta(r,0)=\bfx_\theta(r,2\pi),
\qquad
0< r\le R,
\label{closure0}
\ee
ensuring that $\bfx$ smoothly maps each concentric circle belonging to the disk of radius $R$ into a closed curve on $\calS$.

The area of a surface element of $\calS$ spanned by $\text{d}r$ and $\text{d}\theta$ is simply $|\bfx_r\dr\ctimes\bfx_{\theta}\dtheta|
=|\bfx_r\ctimes\bfx_{\theta}|\drdtheta$; thus,
\be
\int_{\calS}\sigma
=\aint\sigma|\bfx_r\ctimes\bfx_{\theta}|\drdtheta.
\label{4}
\ee

Consistent with \eqref{parametrization}, $\calC=\partial\calS$ can be parameterized according to
\be
\calC=\{\bfx\in\bbR^3:\bfx=\bfx(R,\theta),0\le\theta\le2\pi\},
\label{5}
\ee
where, to embody the assumption that the bounding loop spanned by the soap film is inextensible, $\bfx$ must satisfy the constraint
\be
|\bfx_{\theta}|_{r\eql R}=R.
\label{6}
\ee

As a consequence of the parametrization \eqref{5}, the curvature $\kappa$ of $\calC$ may be viewed as a function of $\theta$. Indeed, by \eqref{6} and the identity $|\bfa\ctimes\bfb|^2=|\bfa|^2|\bfb|^2-(\bfa\cdot\bfb)^2$, it follows that
\be
\left.\kappa=\frac{|\bfx_\theta\times\bfx_{\theta\theta}|}
{|\bfx_{\theta}|^3}\bigg|_{r\eql R}
=\frac{\sqrt{|\bfx_\theta|^2|\bfx_{\theta\theta}|^2-\half(\bfx_\theta\cdot\bfx_\theta)_\theta}}
{|\bfx_\theta|^3}\right|_{r\eql R}
=\frac{|\bfx_{\theta\theta}|_{r\eql R}}{R^2};
\label{7}
\ee
thus, bearing in mind that $R\dtheta$ is the element of arclength corresponding to the parametrization \eqref{5} of $\calC$,
\be
\int_{\calC}\half a\kappa^2
=\lint\frac{a|\bfx_{\theta \theta}|^2}{2R^3}\bigg|_{r\eql R}\dtheta.
\label{8}
\ee

In view of the parameterized versions \eqref{4} and \eqref{8} of the areal and lineal contributions to the net free-energy, \eqref{energy0} admits a representation,
\be
F[\bfx]=\aint\sigma|\bfx_r\ctimes\bfx_{\theta}|\drdtheta
+\lint\frac{a|\bfx_{\theta \theta}|^2}{2R^3}\bigg|_{r\eql R}\dtheta,
\label{energy}
\ee
as a functional of $\bfx$. Given an orthogonal tensor $\bfQ$ and a vector $\bfd$, it is easily verified that
\be
F[\bfQ\bfx+\bfd]=F[\bfx]
\label{11}
\ee
and, thus, that the net free-energy is invariant under rigid transformations.

The parametrization \eqref{parametrization} results in significant analytical advantages. With it, the net free-energy of the system comprised a soap film bounded by an inextensible elastic loop is a functional of a mapping, namely $\bfx$, instead of a set, namely $\calS$. This makes it possible to apply without difficulty the complete apparatus of variational methods for deriving equations of equilibrium and stability conditions, as presented in Section 4. 

\section{Scaling}
\label{scaling}

For the simple choice
\be
\bfx=r\mskip1mu\bfe,
\label{groundstate}
\ee
which represents a flat disk-like surface bounded by circular loop of radius $R$ and can thus be thought of as a ground state, \eqref{energy} specializes to yield a reference value, %
\be
F[r\mskip1mu\bfe]=\bigg(1+\frac{R^3\sigma}{a}\bigg)\frac{\pi a}{R},
\label{referenceF}
\ee
of the net free-energy $F$. The dimensionless parameter
\be
\nu=\frac{R^3\sigma}{a}>0
\label{nu}
\ee
entering \eqref{referenceF} emerges as a natural measure of the relative importance of the two contributions to $F$; specifically, $\nu$ is the ratio of the areal free-energy $\pi R^2\sigma$ to the lineal bending-energy $\pi a/R$ for a ground state of the form \eqref{groundstate} describing a circular disk of radius $R$.

Instead of $\nu$, \cite{gm} work with the dimensionless parameter $\gamma$ defined in \eqref{gammaintro} and, by \eqref{nu}, related to $\nu$ by
\be
\gamma=16\pi^3\nu.
\label{gamma}
\ee

\section{Energy stability criterion}
\label{energystablity}

By the energy stability criterion, as treated by \cite{ericksen}, a stable equilibrium of the system consisting of the bounding loop and the spanning surface is given by a parametrization $\bfx$ that minimizes the net free-energy $F$ subject to the inextensibility constraint \eqref{6}. In applying this criterion, recall that $\bfx$ is assumed to be three-times continuously differentiable on the closed disk of radius $R$.

\subsection{First variation condition}

To compute the first variation $\delta F$ of the functional $F$ defined in \eqref{energy}, consider a smooth variation
\be
\delta\bfx=\bfu
\label{bfu}
\ee
of the parametrization $\bfx$. Further, define a unit normal $\bfn$ to the spanning surface $\calS$ by
\be
\bfn=\frac{\bfx_r\ctimes\bfx_\theta}{|\bfx_r\ctimes\bfx_\theta|}.
\label{14}
\ee

It is convenient to examine separately the areal and lineal contributions \eqref{4} and \eqref{8} to $F$. First, varying \eqref{4} and using integration by parts yields
\begin{align}
\delta\aint\sigma|\bfx_r\ctimes\bfx_{\theta}|\drdtheta
&=\aint\sigma[(\bfx_\theta\ctimes\bfn)\cdot\bfu_r+(\bfn\ctimes\bfx_r)\cdot\bfu_\theta]
\drdtheta \;
\notag\\[4pt]
&=-\aint\sigma[(\bfx_\theta\ctimes\bfn)_r+(\bfn\ctimes\bfx_r)_\theta]\cdot\bfu
\drdtheta
\notag\\[4pt]
&\quad\qquad+\lint\sigma[(\bfx_\theta\times\bfn)\cdot\bfu]_{r\eql R}\dtheta
\notag\\[4pt]
&=-\aint\sigma(\bfx_\theta\ctimes\bfn_r+\bfn_\theta\ctimes\bfx_r)\cdot\bfu
\drdtheta
\notag\\[4pt]
&\quad\qquad+\lint\sigma[(\bfx_\theta\times\bfn)\cdot\bfu]_{r\eql R}\dtheta.
\label{arealvar}
\end{align}
Next, varying \eqref{8} and using integration by parts yields
\begin{align}
\delta\lint\frac{a|\bfx_{\theta \theta}|^2}{2R^3}\bigg|_{r\eql R}\dtheta
&=\lint\frac{a[\bfx_{\theta\theta}\cdot\bfu_{\theta\theta}]_{r\eql R}}{R^3}\dtheta
\notag\\[4pt]
&=-\lint\frac{a[\bfx_{\theta\theta\theta}\cdot\bfu_\theta]_{r\eql R}}{R^3}\dtheta
\label{linealvar0}
\end{align}
However, in view of the constraint \eqref{6}, $\bfx_\theta$ and $\bfu_\theta$ must obey
\be
[\bfx_\theta\cdot\bfu_\theta]_{r\eql R}=0,
\label{19}
\ee
from which it follows that
\be
[\bfx_{\theta\theta\theta}\cdot\bfu_\theta]_{r\eql R}
=[(\bfx_{\theta\theta\theta}-\lambda\mskip1.5mu\bfx_\theta)\cdot\bfu_\theta]_{r\eql R},
\ee
where $\lambda$ is a dimensionless, scalar-valued Lagrange multiplier needed to ensure satisfaction of the inextensibility constraint \eqref{6}.

Further, since, by \eqref{closure0},
\be
[(\bfx_{\theta\theta\theta}-\lambda\mskip1.5mu\bfx_\theta)\cdot\bfu]_{(r,\theta)\eql(R,2\pi)}
=[(\bfx_{\theta\theta\theta}-\lambda\mskip1.5mu\bfx_\theta)\cdot\bfu]_{(r,\theta)\eql(R,0)},
\label{linealvarclosure}
\ee
\eqref{linealvar0} becomes
\begin{align}
\delta\lint\frac{a|\bfx_{\theta \theta}|^2}{2R^3}\bigg|_{r\eql R}\dtheta
&=-\lint\frac{a[(\bfx_{\theta\theta\theta}-\lambda\mskip1.5mu\bfx_\theta)\cdot\bfu_\theta]_{r\eql R}}{R^3}\dtheta
\notag\\[4pt]
&=\lint\frac{a[(\bfx_{\theta\theta\theta}-\lambda\mskip1.5mu\bfx_\theta)_\theta\cdot\bfu]_{r\eql R}}{R^3}\dtheta,
\label{linealvar}
\end{align}
where integration by parts has again been employed.

Finally, the areal and lineal variations \eqref{arealvar} and \eqref{linealvar} combine to yield
\begin{multline}
\delta F=-\aint\sigma(\bfx_\theta\ctimes\bfn_r+\bfn_\theta\ctimes\bfx_r)\cdot\bfu\drdtheta
\\
+\lint\frac{a[(\nu\mskip1mu\bfx_\theta\ctimes\bfn+
(\bfx_{\theta\theta\theta}
-\lambda\mskip1.5mu\bfx_\theta)_\theta)\cdot\bfu]_{r\eql R}}{R^3}\dtheta,
\label{variation}
\end{multline}
where the definition \eqref{nu} of the dimensionless parameter $\nu$ has been invoked.

\subsection{Equilibrium conditions}

Bearing in mind \eqref{variation} and the fundamental theorem of the calculus of variations, the first-variation condition $\delta F=0$ delivers two equilibrium conditions necessary for the parametrization $\bfx$ to extremize $F$. Specifically, requiring that $\delta F=0$ for all variations $\bfu=\delta\bfx$ compactly supported on the interior of the disk of radius $R$ gives the areal equilibrium condition
\be
\bfx_\theta\ctimes\bfn_r+\bfn_\theta\ctimes\bfx_r={\bf0},
\label{16dim}
\ee
where a common factor of the positive parameter $\sigma$ has been dropped from each term. Further, requiring that $\delta F=0$ for all variations $\bfu=\delta\bfx$ with compact support including a segment of the circle of radius $R$ gives the lineal equilibrium condition
\be
[\nu\mskip1mu\bfx_\theta\ctimes\bfn+
(\bfx_{\theta\theta\theta}
-\lambda\mskip1.75mu\bfx_\theta)_\theta]_{r\eql R}=\bf0.
\label{17dim}
\ee
The areal and linear equilibrium conditions \eqref{16dim} and \eqref{17dim} are to be supplemented by the inextensibility constraint
\be
|\bfx_{\theta}|_{r\eql R}=R
\label{12bis}
\ee
and the closure conditions
\be
\bfx(r,0)=\bfx(r,2\pi)
\qquad\text{and}\qquad
\bfx_\theta(r,0)=\bfx_\theta(r,2\pi),
\qquad
0< r\le R.
\label{closure0bis}
\ee
Of \eqref{16dim} and \eqref{17dim}, it is only \eqref{17dim} that involves $\nu$ and, hence, that embodies the competition between the the areal and lineal contributions to the net free-energy $F$.

Geometric and kinetic interpretations of the equilibrium conditions \eqref{16dim} and \eqref{17dim} are provided in Sections~\ref{geometriccontent} and \ref{kineticcontent}, respectively. The unusual nature of the boundary-value problem \eqref{16dim}--\eqref{closure0bis} and some attendant mathematical challenges are discussed in Section~\ref{challenges}.

\subsection{Second variation condition}

A parametrization $\bfx$ satisfying the equilibrium conditions \eqref{16dim} and \eqref{17dim} is stable only if it satisfies the second-variation condition $\delta^2\mskip-2mu F\ge0$. To determine an explicit expression for $\delta^2\mskip-2mu F$, consider once again a smooth variation $\delta\bfx=\bfu$ of the parametrization $\bfx$. Following the approach leading to the first variation, consider separately the areal and lineal contributions \eqref{4} and \eqref{8} to $F$. To begin, the equality on the first line of \eqref{arealvar} may be written as
\be
\delta\aint\sigma|\bfx_r\ctimes\bfx_{\theta}|\drdtheta
=\aint\sigma\bfn\cdot(\bfu_r\ctimes\bfx_\theta+\bfx_r\ctimes\bfu_\theta)\drdtheta,
\label{delta2aint0}
\ee
from which it follows that
\be
\delta^2\aint\sigma|\bfx_r\ctimes\bfx_{\theta}|\drdtheta
=\aint\sigma
(\delta\bfn\cdot\delta(\bfx_r\ctimes\bfx_{\theta})
+\bfn\cdot\delta^2(\bfx_r\ctimes\bfx_{\theta}))\drdtheta.
\label{2vaint}
\ee
Toward simplifying \eqref{2vaint}, notice that
\be
|\bfx_r\ctimes\bfx_{\theta}|\delta\bfn
=(\idem-\bfn\otimes\bfn)\delta(\bfx_r\ctimes\bfx_{\theta})
=(\idem-\bfn\otimes\bfn)(\bfu_r\ctimes\bfx_\theta+\bfx_r\ctimes\bfu_\theta),
\ee
where $\idem$ denotes the identity tensor.
Thus, on introducing the perpendicular projector
\be
\bfP=\idem-\bfn\otimes\bfn
\label{projector}
\ee
onto the tangent plane of the spanning surface, it follows that
\be
\delta\bfn\cdot\delta(\bfx_r\ctimes\bfx_{\theta})
=\frac{|\bfP(\bfu_r\ctimes\bfx_\theta+\bfx_r\ctimes\bfu_\theta)|^2}{|\bfx_r\ctimes\bfx_{\theta}|}.
\label{2ndvari1}
\ee
Further, with the areal equilibrium condition \eqref{16dim} taken into consideration,
\begin{align}
\bfn\cdot\delta^2(\bfx_r\ctimes\bfx_{\theta})
&=\bfn\cdot(\delta\bfu_r\times\bfx_\theta+2\bfu_r\times\bfu_\theta+\bfx_r\times\delta\bfu_\theta)
\notag
\\[4pt]
&=(\bfx_\theta\times\bfn)\cdot\delta\bfu_r+2\bfn\cdot(\bfu_r\times\bfu_\theta)
+(\bfn\times\bfx_r)\cdot\delta\bfu_\theta
\notag
\\[4pt]
&=((\bfx_\theta\times\bfn)\cdot\delta\bfu)_r+2\bfn\cdot(\bfu_r\times\bfu_\theta)
+((\bfn\times\bfx_r)\cdot\delta\bfu)_\theta
\notag
\\
&\qquad-((\bfx_\theta\times\bfn)_r+(\bfn\times\bfx_r)_\theta)\cdot\delta\bfu
\notag
\\[4pt]
&=((\bfx_\theta\times\bfn)\cdot\delta\bfu)_r+2\bfn\cdot(\bfu_r\times\bfu_\theta)
+((\bfn\times\bfx_r)\cdot\delta\bfu)_\theta
\notag
\\
&\qquad
-(\bfx_\theta\times\bfn_r+\bfn_\theta\times\bfx_r
+\bfx_{r\theta}\times\bfn+\bfn\times\bfx_{r\theta})\cdot\delta\bfu
\notag
\\[4pt]
&=((\bfx_\theta\times\bfn)\cdot\delta\bfu)_r+2\bfn\cdot(\bfu_r\times\bfu_\theta)
+((\bfn\times\bfx_r)\cdot\delta\bfu)_\theta.
\label{2ndvari2}
\end{align}
On inserting \eqref{2ndvari1} and \eqref{2ndvari2} in \eqref{2vaint}, using integration by parts, and invoking the definition \eqref{nu} of $\nu$, it follows that
\begin{multline}
\delta^2\aint\sigma|\bfx_r\ctimes\bfx_{\theta}|\drdtheta
=\aint\sigma\bigg(\frac{|\bfP(\bfu_r\ctimes\bfx_\theta+\bfx_r\ctimes\bfu_\theta)|^2}
{|\bfx_r\ctimes\bfx_{\theta}|}
+2\bfn\cdot(\bfu_r\times\bfu_\theta)
\bigg)\drdtheta
\\[4pt]
+\lint\frac{a[\nu(\bfx_\theta\times\bfn)\cdot\delta\bfu]_{r\eql R}}{R^3}\dtheta.
\label{delta2aint}
\end{multline}
Next, in view of the equality on the first line of \eqref{linealvar0}, the second variation of the lineal contribution \eqref{8} to $F$ is given by
\begin{align}
\delta^2\lint\frac{a|\bfx_{\theta \theta}|^2}{2R^3}\bigg|_{r\eql R}\dtheta
&=\lint\frac{a[|\bfu_{\theta\theta}|^2
+\bfx_{\theta\theta}\cdot \delta \bfu_{\theta\theta}]_{r\eql R}}{R^3}\dtheta
\notag
\\[4pt]
&=\lint\frac{a[|\bfu_{\theta\theta}|^2
+\bfx_{\theta\theta\theta\theta}\cdot \delta \bfu]_{r\eql R}}{R^3}\dtheta.
\label{delta2lint}
\end{align}

By \eqref{delta2aint}, \eqref{delta2lint}, and the lineal equilibrium condition \eqref{17dim}, the second variation condition $\delta^2\mskip-2mu F\ge0$ can be expressed as
\begin{multline}
\delta^2\mskip-2mu F
=\aint\sigma\bigg(\frac{|\bfP(\bfu_r\ctimes\bfx_\theta+\bfx_r\ctimes\bfu_\theta)|^2}
{|\bfx_r\ctimes\bfx_{\theta}|}
+2\bfn\cdot(\bfu_r\times\bfu_\theta)
\bigg)\drdtheta
\\[4pt]
+\lint\frac{a[|\bfu_{\theta\theta}|^2
+(\lambda\bfx_{\theta})_\theta\cdot\delta\bfu]_{r\eql R}}{R^3}\dtheta\ge0.
\label{delta2E0}
\end{multline}
However,  on differentiating the constraint \eqref{6} to yield the identity
\be
[\bfx_\theta\cdot\delta\bfu_\theta+|\bfu_\theta|^2]_{r\eql R}=0
\label{auxconstaint}
\ee
and using integration by parts, \eqref{delta2E0} simplifies to
\begin{multline}
\delta^2\mskip-2mu F=
\aint\sigma\bigg(\frac{|\bfP(\bfu_r\ctimes\bfx_\theta+\bfx_r\ctimes\bfu_\theta)|^2}{|\bfx_r\ctimes\bfx_\theta|}
+2\bfn\cdot(\bfu_r\ctimes\bfu_\theta)\bigg)\drdtheta
\\[4pt]
+\lint\frac{a(|\bfu_{\theta\theta}|^2+\lambda|\bfu_{\theta}|^2)|_{r\eql R}}{R^3}\dtheta\ge0.
\label{18dim}
\end{multline}
Although the lineal contribution to $\delta^2\mskip-2mu F$ is independent of the parametrization $\bfx$, it depends on the Lagrange multiplier $\lambda$.

\section{Geometric content of the equilibrium conditions}
\label{geometriccontent}

\subsection{Preliminary definitions}

Aside from the unit normal $\bfn$ and projector $\bfP$ defined in \eqref{14} and \eqref{projector}, the only quantity needed to explore the geometric content of areal equilibrium condition \eqref{16dim} is the mean curvature
\be
H=\frac{\bfn\cdot(|\bfx_\theta|^2\bfx_{rr}
-2(\bfx_r\cdot\bfx_\theta)\bfx_{r\theta}
+|\bfx_r|^2\bfx_{\theta\theta})}
{2|\bfx_r\ctimes\bfx_\theta|^2}
\label{H}
\ee
of the spanning surface $\calS$. With the identities
\be
\bfn\cdot(\bfx_\theta\ctimes\bfn_r+\bfn_\theta\ctimes\bfx_r)
=(\bfn_r\ctimes\bfn)\cdot\bfx_\theta+(\bfn\ctimes\bfn_\theta)\cdot\bfx_r.
\ee
and
\be
\left.
\begin{split}
\bfn_r\ctimes\bfn
&=\bigg(\frac{\bfx_r\ctimes\bfx_\theta}{|\bfx_r\ctimes\bfx_\theta|}
\bigg)_{\mskip-4mu r}\ctimes\bfn
=\frac{(\bfn\cdot\bfx_{rr})\bfx_\theta
-(\bfn\cdot\bfx_{r\theta})\bfx_r}
{|\bfx_r\ctimes\bfx_\theta|},
\\[4pt]
\bfn\ctimes\bfn_\theta
&=\bfn\ctimes
\bigg(\frac{\bfx_r\ctimes\bfx_\theta}{|\bfx_r\ctimes\bfx_\theta|}
\bigg)_{\mskip-4mu\theta}
=\frac{(\bfn\cdot\bfx_{\theta\theta})\bfx_r
-(\bfn\cdot\bfx_{r\theta})\bfx_\theta}
{|\bfx_r\ctimes\bfx_\theta|},
\end{split}
\,\right\}
\ee
\eqref{H} yields a useful alternative expression
\be
H=\frac{\bfn\cdot(\bfx_\theta\ctimes\bfn_r+\bfn_\theta\ctimes\bfx_r)}
{2|\bfx_r\ctimes\bfx_{\theta}|},
\label{Halt}
\ee
for the mean curvature of $\calS$.

To explore the geometric content of the lineal equilibrium condition \eqref{17dim}, it is convenient to convert from the parametrization of $\calC$ embodied by \eqref{5} to an arclength parametrization. To achieve this, introduce the arclength
\be
s=R\mskip1mu\theta,
\qquad
0\le\theta\le2\pi,
\label{arclength}
\ee
and define $\bar\bfx$ such that
\be
\bar\bfx(R\mskip1mu\theta)=\bfx(R,\theta)
\qquad
0\le\theta\le2\pi,
\label{barx}
\ee
in which case \eqref{5} becomes
\be
\calC=\{\bfx\in\bbR^3:\bfx=\bar\bfx(s),0\le s\le L\}.
\label{Carclengthparam}
\ee
Since, by \eqref{barx}, $\bfx_\theta|_{r\eql R}=R\bar\bfx_s$, the stipulation \eqref{12bis} that $\calC$ be inextensible becomes
\be
|\bar\bfx_s|=1,
\label{barconstraint}
\ee
which confirms that $\bar\bfx$ does indeed provide an arclength parametrization of $\calC$. Further, the definition \eqref{7} of the curvature $\kappa$ of $\calC$ becomes
\be
\kappa=|\bar\bfx_{ss}|.
\label{barcurvature}
\ee

Aside from $\kappa$, the quantities needed to explore the geometric content of the lineal equilibrium condition \eqref{17dim} are the torsion $\tau$ of $\calC$, defined consistent with
\be
\kappa^2\tau=(\bar\bfx_s\ctimes\bar\bfx_{ss})\cdot\bar\bfx_{sss},
\label{tau}
\ee
and the Frenet frame $\{\bft,\bfp,\bfb\}$ of $\calC$, with tangent, normal, and binormal elements defined consistent with
\be
\bft=\bar\bfx_s,
\qquad
\kappa\mskip1mu\bfp=\bar\bfx_{ss},
\qquad\text{and}\qquad
\kappa\mskip1mu\bfb=\bar\bfx_s\ctimes\bar\bfx_{ss}.
\label{Ffdefs}
\ee
In working with these objects, the Frenet--Serret relations,
\be
\bft_s=\kappa\mskip1mu\bfp,
\qquad
\bfp_s=-\kappa\mskip1mu\bft+\tau\mskip1mu\bfb,
\qquad\text{and}\qquad
\bfb_s=-\tau\bfp,
\label{FSrels}
\ee
are indispensable.

It is also useful to express the restriction $\bfn|_{r\eql R}$ of the unit normal $\bfn$ along the boundary $\partial\calS$ of the spanning surface in terms of arclength. To achieve this, define $\bar\bfn$ such that
\be
\bar\bfn(R\mskip1mu\theta)=\bfn(R,\theta),
\qquad
0\le\theta\le2\pi.
\label{barn}
\ee
Then, following \cite{gm}, $\bar\bfn$ can be expressed in terms of the ``contact angle" $\vartheta$, the principal normal $\bfp$, and the principal binormal $\bfb$ belonging to the Frenet frame of $\calC$ via
\be
\bar\bfn=(\cos\vartheta)\mskip1mu\bfp+(\sin\vartheta)\mskip1mu\bfb.
\label{b3}
\ee

\subsection{Areal equilibrium condition}

It is evident from \eqref{Halt} that the areal equilibrium condition \eqref{16dim} implies that the mean curvature vanishes. Conversely, since, by \eqref{14}, the dot products $\bfn\cdot\bfn_r$, $\bfn\cdot\bfn_\theta$, $\bfn\cdot\bfx_r$, and $\bfn\cdot\bfx_\theta$ must all vanish, each of  $\bfx_\theta\ctimes\bfn_r$ and $\bfn_\theta\ctimes\bfx_r$ must either be a multiple of $\bfn$ or vanish. Hence,
\be
\bfn\cdot(\bfx_\theta\ctimes\bfn_r+\bfn_\theta\ctimes\bfx_r)=0
\label{arealscal}
\ee
implies \eqref{16dim}. The areal equilibrium condition \eqref{16dim} is satisfied if and only if $\calS$ has zero mean curvature.

Alternatively, the foregoing calculations show that \eqref{16dim} is equivalent to the requirement
\be
H\bfn=\bf0
\label{vectorpde}
\ee
that the vector mean curvature of $\calS$ vanishes.

\subsection{Lineal equilibrium condition}

Using the definitions \eqref{nu}, \eqref{barx}, \eqref{Ffdefs}, and \eqref{barn} of $\nu$, $\bar\bfx$, $\{\bft,\bfp,\bfn\}$, and $\bar\bfn$ in the lineal equilibrium condition \eqref{17dim} yields
\be
\frac{\sigma\mskip1mu\bft\ctimes\bar\bfn}{a}
+\bigg(\bft_{ss}-\frac{\lambda\mskip1mu\bft}{R^2}\bigg)_{\mskip-6mu s}=\bf0,
\label{17dim2}
\ee
which, on invoking the Frenet--Serret relations \eqref{FSrels} and the defining relation \eqref{b3} for the contact angle $\vartheta$, is equivalent to
\be
\bigg(\frac{3}{2}\kappa^2+\frac{\lambda}{R^2}\bigg)_{\mskip-6mu s}\mskip1mu\bft
-\bigg(\kappa_{ss}-\kappa^3-\bigg(\tau^2+\frac{\lambda}{R^2}\bigg)\kappa
-\frac{\sigma}{a}\sin\vartheta\bigg)\bfp
-\bigg(2\kappa_s\tau+\kappa\mskip1mu\tau_s+\frac{\sigma}{a}\cos\vartheta\bigg)\bfb
=\bf0.
\label{17t}
\ee
Since the Frenet frame $\{\bft,\bfp,\bfb\}$ of $\calC$ is orthonormal, \eqref{17t} cannot be satisfied unless the coefficients of each term entering \eqref{17t} vanish. Isolating first the term parallel to the tangent $\bft$ of $\calC$, the requirement that its coefficient vanishes implies that the Langrange multiplier $\lambda$ is given by
\be
\lambda=\frac{R^2}{a}\bigg(\beta-\frac{3}{2}a\kappa^2\bigg),
\label{teqn}
\ee
with $\beta$ being a constant with the dimensions of energy per unit length of $\calC$. Notice that $a\mskip1mu\lambda/R^2$ is the reactive force required to ensure satisfaction of the constraint \eqref{12bis} of inextensibility. Focussing next on the term parallel to the principal normal $\bfp$ of $\calC$, bearing in mind \eqref{teqn}, the requirement that its coefficient vanishes yields
\be
\kappa_{ss}+\frac{1}{2}\kappa^3-\bigg(\tau^2+\frac{\beta}{a}\bigg)\kappa-\frac{\sigma}{a}\sin\vartheta=0.
\label{b1}
\ee
Focusing finally on the term parallel to the bionormal $\bfb$ of $\calC$, the requirement that its coefficient vanishes yields
\be
2\kappa_s\tau+\kappa\mskip1mu\tau_s+\frac{\sigma}{a}\cos\vartheta=0.
\label{b2}
\ee
Thus, \eqref{17dim} implies \eqref{teqn}--\eqref{b2}. Conversely, since \eqref{17t} is equivalent to \eqref{17dim}, using \eqref{teqn}--\eqref{b2} in \eqref{17t} implies that \eqref{17dim} holds. The lineal equilibrium condition \eqref{17dim} is therefore equivalent to the requirement that the Lagrange multiplier $\lambda$ be related to the curvature $\kappa$ of $\calC$ by \eqref{teqn} and that the curvature $\kappa$ and torsion $\tau$ of $\calC$ and the contact angle $\vartheta$ must satisfy \eqref{b1} and \eqref{b2}.

Equations \eqref{b1} and \eqref{b2} correspond to (2.8a) and (2.8b) of \cite{gm}, who claim that ``Determining the shape of the soap film then corresponds to solving (2.8) subject to periodicity of the boundary curve.'' However, \eqref{b1} and \eqref{b2} consist of two equations in four unknown variables, namely $\kappa$, $\tau$, $\vartheta$, and $\beta$ (which, however, might be determined by requiring that the net length of the bounding loop be fixed). Indeed, as opposed to working with their (2.8), \cite{gm} employ a numerical strategy based on minimizing a discrete analogue of the net free-energy \eqref{energy0}. 

\section{Kinetic content of the equilibrium conditions}
\label{kineticcontent}

Since they arise from varying the net free-energy $F$, it is to be expected that the areal and lineal equilibrium conditions \eqref{16dim} and \eqref{17dim} contain information concerning the balances of forces and moments. To clarify the kinetic content of \eqref{16dim} and \eqref{17dim}, suppose that $\calS$ is a material surface with a flexible but inextensible boundary $\calC=\partial\calS$ capable of sustaining forces and moments. Let $\bfT$ denote the superficial Cauchy stress tensor, with dimensions of force per unit length, on $\calS$ and let $\bff$ and $\bfm$ denote, respectively, the force and moment on $\calC$.

\subsection{Areal equilibrium condition}

Consider an arbitrary subsurface $\calA$ of $\calS$. Suppose that $\calA\cap\calC$ is empty, so that $\calA$ is interior to $\calS$. The force and moment balances for $\calA$ then read
\be
\int\limits_{\partial\calA}\bfT(\bft\times\bar\bfn)=\bf0
\qquad\text{and}\qquad
\int\limits_{\partial\calA}(\bfx-\bfx_0)\times\bfT(\bft\times\bar\bfn)=\bf0,
\label{fbA0}
\ee
where $\bfx_0$ denotes an arbitrary point in space. By the surface divergence theorem and the arbitrary nature of $\calA$, \eqref{fbA0} is equivalent to pointwise balances
\be
\text{div}_{\scriptscriptstyle\calS}\bfT=\bf0
\qquad\text{and}\qquad
\bfT=\bfT^{\trans},
\label{fbA}
\ee
where $\text{div}_{\scriptscriptstyle\calS}$ is the divergence operator on $\calS$ and $\bfT^{\trans}$ denotes the transpose of $\bfT$.

Suppose that the superficial Cauchy stress $\bfT$ is determined by the particular constitutive relation
\be
\bfT=\sigma\mskip-1.5mu\idem_{\mskip-4mu\scriptscriptstyle\calS}\label{Trel}
\ee
where $\sigma$ is now to be thought of as a constant surface tension and $\idem_{\mskip-4mu\scriptscriptstyle\calS}$ denotes the identity tensor on the tangent plane of $\calS$. Notice that $\idem_{\mskip-4mu\scriptscriptstyle\calS}$ is the restriction of $\bfP$ to the tangent plane of $\cal S$. In view of the identity $\text{div}_{\scriptscriptstyle\calS}\idem_{\mskip-4mu\scriptscriptstyle\calS}=2H\bfn$,\footnote{See, for instance, Proposition 2.1 of \cite{gm75}.} using \eqref{Trel} in the pointwise force balance \eqref{fbA}$_1$ yields $2\mskip1mu\sigma H\bfn=\bf0$ or, equivalently, since $\sigma$ is presumed nonnegative,
\be
H\bfn=\bf0,
\label{arealpointwisefb}
\ee
which is the version of the areal equilibrium condition appearing in \eqref{vectorpde}. Conversely, integrating \eqref{arealpointwisefb} over $\calA$, and using the surface divergence theorem yields
\be
\int\limits_{\partial\calA}\sigma(\bft\times\bar\bfn)=\bf0,
\ee
which is the particular form of the force balance \eqref{fbA0} for $\calA$ when the superficial Cauchy stress $\bfT$ is of the isotropic tensile form \eqref{Trel}. It therefore follows that, in conjunction with particular constitutive prescription \eqref{Trel} for $\bfT$, the areal equilibrium condition \eqref{16dim} is a pointwise expression of force balance on $\calS$.

Since, in equilibrium, the surface tension of a single-component fluid coincides its free-energy density, it is expected that the relation \eqref{Trel} for $\bfT$ should derive from a free-energy density. To clarify this, it is convenient to treat the parametrization $\bfx(r,\theta)$ of $\cal S$ as a deformation of an elastic surface identified with a disk of radius $R$ to a surface $\cal S$. The value $W(r,\theta)$ of the superficial free-energy density $W$, measured per unit area of the disk, at a point $(r,\theta)$ in the interior of that disk is then determined by a relation of the form
\be
W(r,\theta)=\hat{W}(\bflambda^r(r,\theta),\bflambda^\theta(r,\theta)),
\ee
where $\bflambda^r$ and $\bflambda^\theta$ are defined by
\be
\bflambda^r(r,\theta)=\bfx_r(r,\theta) \qquad\text{and}\qquad \bflambda^{\theta}(r,\theta)=\frac{1}{r}\bfx_{\theta}(r,\theta).
\ee
Moreover, the corresponding value $\bfT(r,\theta)$ of the superficial Cauchy stress then has the form
\be
\bfT(r,\theta)=\frac{\bfsigma^r(r,\theta)\otimes\bflambda^r(r,\theta)+\bfsigma^\theta(r,\theta)\otimes\bflambda^\theta(r,\theta)}
{|\bflambda^r(r,\theta)\ctimes\bflambda^\theta(r,\theta)|},
\label{surfacestress}
\ee
where $\bfsigma^r$ and $\bfsigma^\theta$ are determined from $\hat{W}$ by
\be
\bfsigma^r=\frac{\partial \hat{W}(\bflambda^r,\bflambda^\theta)}{\partial\bflambda^r}
\qquad\text{and}\qquad
\bfsigma^\theta=\frac{\partial \hat{W}(\bflambda^r,\bflambda^\theta)}{\partial\bflambda^\theta}.
\ee
Suppose, now, that the superficial free-energy density is proportional to the area ratio and, thus, that the response function $\hat{W}$ has the particular form
\be
\hat{W}(\bflambda^r,\bflambda^\theta)
=\sigma|\bflambda^r\ctimes\bflambda^\theta|,
\label{strainenergy}
\ee
with $\sigma$ being a constant measure of free energy per unit area. Direct calculations then yield
\be
\bfT=\sigma\bfM(\bflambda^r,\bflambda^\theta),
\label{surfacestress}
\ee
with
\be
\bfM(\bflambda^r,\bflambda^\theta)=
\frac{|\bflambda^\theta|^2\bflambda^r\otimes\bflambda^r
-(\bflambda^r\cdot\bflambda^\theta)(\bflambda^r\otimes\bflambda^\theta+\bflambda^\theta\otimes\bflambda^r)
+|\bflambda^r|^2\bflambda^\theta\otimes\bflambda^\theta}
{|\bflambda^r\times\bflambda^\theta|^2}.
\label{Mdef}
\ee
It is evident from \eqref{Mdef} that
\be
\bfM(\bflambda^r,\bflambda^\theta)\bflambda^r=\bflambda^r
\qquad\text{and}\quad
\bfM(\bflambda^r,\bflambda^\theta)\bflambda^\theta=\bflambda^\theta,
\label{Mreq1}
\ee
for all $(\bflambda^r,\bflambda^\theta)$, which, together, imply that $\bfM(\bflambda^r,\bflambda^\theta)$ is the superficial identity tensor $\idem_{\mskip-4mu\scriptscriptstyle\calS}$. The choice \eqref{strainenergy} of the response function $\hat{W}$ determining the superficial free-energy density $W$ therefore yields the particular superficial Cauchy stress \eqref{Trel} for which force balance \eqref{fbA}$_1$ specializes to the areal equilibrium condition \eqref{16dim}.

\subsection{Lineal equilibrium condition}

Consider an arbitrary segment $\calL$ of $\calC$ and let $\bar\bfx_1$ and $\bar\bfx_2$ denote the points at which $\calL$ begins and ends granted that $\calC$ is traversed in the direction of increasing $\theta$. The force and moment balances for $\calL$ then read
\be
-\int\limits_{\calL}\bfT(\bft\times\bar\bfn)+\bff|_{\scriptscriptstyle\partial\calL}=\bf0
\label{fbL0}
\ee
and
\be
-\int\limits_{\calL}(\bar\bfx-\bar\bfx_0)\times\bfT(\bft\times\bar\bfn)
+[(\bar\bfx-\bar\bfx_0)\times\bff+\bfm]_{\scriptscriptstyle\partial\calL}=\bf0,
\label{mbL0}
\ee
where, for example,
\be
\bff|_{\scriptscriptstyle\partial\calL}=\bff(\bar\bfx_2)-\bff(\bar\bfx_1).
\ee
By the fundamental theorem of line integration, the arbitrary nature of $\calL$, and \eqref{Ffdefs}$_1$, \eqref{fbL0} and \eqref{mbL0} are equivalent to pointwise balances
\be
-\bfT(\bft\times\bar\bfn)+\bff_s={\bf0}
\qquad\text{and}\qquad
\bft\times\mskip-2mu\bff+\bfm_s=\bf0.
\label{fbL}
\ee

Granted that $\calC$ is an inextensible elastic rod with lineal free-energy density, measured per unit length,
\be
w=\hat{w}(\kappa)=\half a\kappa^2,
\label{arealw}
\ee
the moment $\bfm$ is determined by the constitutive relation
\be
\bfm=\frac{\text{d}\hat{w}(\kappa)}{\text{d}\kappa}\bfb=a\kappa\bfb.
\label{mcl}
\ee
Since, by the Frenet--Serret relations \eqref{FSrels},
\be
(\kappa\bfb)_s=\kappa_s\bfb-\kappa\tau\bfp=\bft\times(\kappa_s\bfp+\kappa\tau\bfb),
\ee
\eqref{mcl} yields
\be
\bft\times\mskip-2mu\bff+\bfm_s=\bft\times(\bff+a(\kappa_s\bfp+\kappa\tau\bfb))
\label{msubs}
\ee
which in view of the pointwise moment balance \eqref{fbL}$_2$ ensures the existence of a constitutively indeterminate scalar field $\mu$ such that $\bff+a(\kappa_s\bfp+\kappa\tau\bfb)=\mu\mskip0.25mu\bft$ or, equivalently, that the force $\bff$ admits a representation of the form
\be
\bff=\mu\mskip0.5mu\bft-a(\kappa_s\bfp+\kappa\tau\bfb)
=(\mu-a\kappa^2)\bft-a(\kappa\bfp)_s.
\label{frel}
\ee
Using the constitutive relation \eqref{Trel} for $\bfT$ and \eqref{frel} in the pointwise force balance \eqref{fbL}$_1$ yields
\be
-\sigma(\bft\times\bar\bfn)+(\mu-a\kappa^2)_s\bft+(\mu-a\kappa^2)\bft_s-a(\kappa\bfp)_{ss}=\bf0.
\label{linealpointwisefb0}
\ee
On invoking both the Frenet--Serret relations \eqref{FSrels} and the defining relation \eqref{b3} for the contact angle and dividing each resulting term by $a>0$, \eqref{linealpointwisefb0} yields
\be
\bigg(\frac{\mu}{a}+\frac{1}{2}\kappa^2\bigg)_{\mskip-6mu s}\mskip1mu\bft
-\bigg(\kappa_{ss}-\tau^2\kappa-\frac{\mu}{a}\kappa
-\frac{\sigma}{a}\sin\vartheta\bigg)\bfp
-\bigg(2\kappa_s\tau+\kappa\mskip1mu\tau_s+\frac{\sigma}{a}\cos\vartheta\bigg)\bfb
=\bf0,
\label{linealpointwisefb}
\ee
which, with the additional identification
\be
\mu=\frac{a\lambda}{R^2}+a\kappa^2,
\label{mulambdarel}
\ee
is the lineal equilibrium condition appearing in \eqref{17t}. Conversely, using \eqref{mulambdarel} in \eqref{frel}, integrating the result over $\calL$, and using the fundamental theorem of line integration yields
\be
-\int\limits_{\calL}\sigma(\bft\ctimes\bar\bfn)
+a\bigg(\frac{\lambda}{R^2}\bft-(\kappa\bfp)_s\bigg)\bigg|_{\scriptscriptstyle\partial\calL}=\bf0,
\ee
which is the particular form of the force balance \eqref{fbL0} for $\calL$ when $\bfT$ is given by \eqref{Trel} and $\bff$ is given by \eqref{frel}. In conjunction with particular constitutive prescription \eqref{Trel} for $\bfT$ and the representation \eqref{frel} of $\bff$, the lineal equilibrium condition \eqref{17dim} is thus a pointwise expression of force balance on $\calC$.

\subsection{Moment balance}

In view of the symmetry of the superficial identity tensor $\idem_\calS$, the choice \eqref{Trel} of the superficial Cauchy stress tensor ensures that moments are balanced pointwise on $\calS$. Moreover, granted that the moment $\bfm$ is determined in accord with \eqref{mcl}, using the representation \eqref{frel} for $\bff$ ensures that the moment balance \eqref{fbL}$_2$ is satisfied on $\calC$. The requirement of moment balance is therefore superfluous in the present work. This would no longer be so in the presence of couples that would accompany endowing $\calS$ with bending elasticity or modeling $\calC$ with a general director theory.

%

\section{Features of the equilibrium boundary-value problem}
\label{challenges}

In view of the definition \eqref{14} of $\bfn$, the areal equilibrium condition \eqref{16dim} is a second-order partial-differential equation. However, the lineal equilibrium condition \eqref{17dim}, which serves as a boundary condition for \eqref{16dim}, is a fourth-order ordinary-differential equation. In this sense, the boundary-value problem comprised by the equilibrium conditions \eqref{16dim} and \eqref{17dim}, the constraint \eqref{12bis} of inextensibility, and the closure conditions \eqref{closure0bis} is highly unconventional. This feature of the problem is illustrated in the linear bifurcation analysis presented in Section~\ref{bifurcation}; there, a consideration of out-of-plane bifurcations from the trivial solution leads to the Laplace equation subject to a fourth-order boundary condition.

Existence results obtained by \cite{by} on the basis of variational arguments seem to support that, provided $\calC$ has no points of self contact, suggest that \eqref{16dim}--\eqref{closure0bis} might constitute a well-posed boundary-value problem for the parametrization $\bfx$ and the Lagrange multiplier $\lambda$ (which are defined on the disk of radius $R$ and its boundary, respectively). However, if self-contact is allowed, as seems necessary in view of the experimental and numerical results of \cite{gm}, issues of existence are likely to become more delicate. In this case, it would be necessary to introduce suitable constraints at points of self-contact, as in the works of \cite{j}, \cite{dlm}, \cite{cs}, and others on the self-contact of supercoiled DNA molecules.

Whereas a solution to the vectorial boundary-value problem \eqref{16dim}--\eqref{closure0bis} would determine the parametrization $\bfx$ of a surface $\calS$ with boundary $\calC=\partial\calS$, together with the Lagrange multiplier $\lambda$, the geometric conditions \eqref{vectorpde}, \eqref{b1}, and \eqref{b2} on the mean curvature $H$ of $\calS$ and the curvature $\kappa$, torsion $\tau$, and contact angle $\vartheta$ of $\calC$ provide no more than a system of constraints on a class of bounded zero mean-curvature surfaces bounded by curves of unknown shape.

Since \eqref{16dim}--\eqref{12bis} imply \eqref{vectorpde}, \eqref{b1}, and \eqref{b2}, it is also noteworthy that any solution to the boundary-value problem \eqref{16dim}--\eqref{12bis} must satisfy \eqref{vectorpde}, \eqref{b1}, and \eqref{b2}. Granted that $\calC$ as parameterized by $\bfx|_{r\eql R}$ is smooth and closed, the curvature $\kappa$, torsion $\tau$, and contact angle $\vartheta$ must be periodic. If, however, $\kappa$, $\tau$, and $\vartheta$ are periodic, it is not necessarily the case that the curve to which they corresponding is closed. To obtain a closed loop requires the imposition of supplemental restrictions on $\kappa$ and $\tau$. Moreover, having used $\kappa$, $\tau$, and said (as yet unidentified) restrictions to determine a smooth, closed loop $\calC$ with parametrization $\bfx|_{r\eql R}$, there remains the problem of constructing a surface $\calS$ with parametrization $\bfx$ as mean curvature $H=0$. It seems likely that doing so might entail the imposition of supplemental restrictions on $\vartheta$.

Recognizing that \cite{gm} work directly with $\alpha=a/2$, replacing $a$ with $2\alpha$ in \eqref{b1} and \eqref{b2} yields their (2.8\emph{a}) and (2.8\emph{b}). Moreover, \eqref{vectorpde} is their (2.8\emph{c}). Consistent with the discussion in the previous paragraph, \cite{gm} mention that their (2.8\emph{a--c}) should be augmented by an additional compatibility condition, namely their (2.7)---which is a condition involving, among other quantities, ``displacements" along the principal normal and binormal directions of the bounding loop. However, an approach to constructing solutions of their system (2.7)--(2.8\emph{a--c}) seems elusive. Indeed, in the remainder of their paper, \cite{gm} generate results mainly by using numerical computation based on a discretized energy, namely their (4.1).

\section{Nondimensionalization}
\label{nondimensionalization}

To facilitate both analysis and the interpretation of results, it is convenient to introduce a dimensionless radial coordinate
\be
\rho=\frac{r}{R}
\label{9m2}
\ee
and a dimensionless dependent variable
\be
\bfxi(\rho,\theta)=\frac{\bfx(R\rho,\theta)}{R},
\label{9m1}
\ee
defined, with reference to \eqref{parametrization}, on the unit disk. By \eqref{9m2} and \eqref{9m1}, the closure conditions \eqref{closure0} become
\be
\bfxi(\rho,0)=\bfxi(\rho,2\pi)
\qquad\text{and}\qquad
\bfxi_\theta(\rho,0)=\bfxi_\theta(\rho,2\pi),
\qquad
0<\rho\le 1.
\label{closure}
\ee
Using \eqref{9m2}--\eqref{9m1} in \eqref{6} yields a dimensionless inextensibility constraint
\be
|\bfxi_{\theta}|_{\rho\eql1}=1,
\label{12}
\ee
from which it follows that the restriction $\bfxi|_{\rho\eql1}$ of $\bfxi$ to the boundary of the unit disk (namely, the unit circle) provides an arclength parametrization of $\calC$ in terms of the remaining independent variable $\theta$, which serves as a dimensionless measure of arclength.

Next, since the reference value \eqref{referenceF} of $F$ scales with $a/R$, it is natural to normalize the net free-energy by that quantity and, thus, work with the dimensionless free-energy functional $\Phi$ defined according to
\be
\Phi[\bfxi]=\frac{F[R\bfxi]}{a/R}
=\aintdim\nu|\bfxi_\rho\ctimes\bfxi_{\theta}|\drhodtheta
+\lint\half|\bfxi_{\theta \theta}|^2\big|_{\rho\eql1}\dtheta.
\label{9}
\ee
Notice that, for the dimensionless counterpart
\be
\bfxi=\rho\mskip1.5mu\bfe
\label{triv2}
\ee
of the ground state \eqref{groundstate}, the corresponding reference value of the dimensionless net free-energy functional \eqref{referenceF} is simply
\be
\Phi[\rho\mskip1.5mu\bfe]=\pi(1+\nu).
\ee

The dimensionless counterparts of the areal and lineal equilibrium conditions \eqref{16dim} and \eqref{17dim}, which are necessary conditions for the vanishing of the first variation $\delta\Phi$ of the dimensionless energy functional $\Phi$ are, respectively,
\be
\bfxi_{\theta}\ctimes\bfn_{\raisebox{-1.15pt}{$\scriptstyle\rho$}}
+\bfn_{\raisebox{-1.15pt}{$\scriptstyle\theta$}}\ctimes\bfxi_{\rho}={\bf0},
\label{16}
\ee
and
\be
[\nu\mskip1mu\bfxi_{\theta}\ctimes\bfn+
(\bfxi_{\theta\theta\theta}
-\lambda\mskip1.75mu\bfxi_{\theta})_{\raisebox{-1.15pt}{$\scriptstyle\theta$}}]_{\rho\eql1}=\bf0.
\label{17}
\ee
Additionally, the dimensionless counterpart of the second-variation condition \eqref{18dim} is
\begin{multline}
\delta^2\Phi=
\aintdim\nu\bigg(\frac{|\bfP(\bfu_{\raisebox{-1.15pt}{$\scriptstyle\rho$}}\ctimes\bfxi_{\theta}+\bfxi_{\rho}\ctimes\bfu_{\raisebox{-1.15pt}{$\scriptstyle\theta$}})|^2}{|\bfxi_{\rho}\ctimes\bfxi_{\theta}|}+2\bfn\cdot(\bfu_{\rho}\ctimes\bfu_{\raisebox{-1.15pt}{$\scriptstyle\theta$}})\bigg)\drhodtheta
\\[4pt]
+\lint(|\bfu_{\theta\theta}|^2+\lambda|\bfu_{\theta}|^2)|_{\rho\eql1} \dtheta\ge0.
\label{18}
\end{multline}

For brevity, the adjective `dimensionless' is hereafter omitted when referring to the independent and dependent variables introduced in \eqref{9m2} and \eqref{9m1}, the constraint \eqref{12}, and the net free-energy \eqref{9}. Additionally, the surface parameterized by $\bfxi$ and the curve parameterized by the restriction $\bfxi|_{\rho\eql1}$ of $\bfxi$ to the boundary of the unit disk are referred to as the `spanning surface' and the `bounding loop,' respectively.

%
%

\section{Stability of trivial solutions}
\label{stability}

The boundary-value problem consisting of the constraint \eqref{12} and the dimensionless areal and lineal equilibrium conditions \eqref{16} and \eqref{17} admits a trivial solution branch with $\bfxi$ and $\lambda$ given by
\be
\bfxi(\rho,\theta)=\rho\mskip1.5mu\bfe,
\qquad
\lambda=-(1+\nu).
\label{20}
\ee
This solution corresponds to the ground state \eqref{triv2}. The trivial solution branch \eqref{20} depends on the parameter $\nu$, which, according to \eqref{nu}, embodies experimental protocols in which $R$ (and, hence, the length $L=2\pi R$ of the bounding loop) may be varied while holding fixed both the surface tension $\sigma$ and the bending rigidity $a$ or where one or both of $\sigma$ and $a$ may be varied while $R$ is held fixed. It is therefore of interest to determine the stability of both trivial solutions and nontrivial solutions that bifurcate from the trivial solution branch \eqref{20} with variations of $\nu$.

By the energy stability criterion discussed in Section~\ref{energystablity}, an equilibrium solution is stable only if it satisfies the second-variation condition \eqref{18}. If this condition is violated for some $\bfu$, then the associated equilibrium solution is unstable and $\bfu$ is a perturbation that lowers the net free-energy.

To evaluate the second-variation condition \eqref{18} at the trivial solution, let
\be
\bfm=\bfe\ctimes\bfeperp
\label{23}
\ee
and notice from \eqref{20} and \eqref{14} that
\be
\bfxi_{\rho}=\bfe,
\qquad
\bfxi_{\theta}=\rho\mskip1.75mu\bfeperp,
\qquad
\bfxi_{\rho}\ctimes\bfxi_{\theta}=\rho\mskip1.75mu\bfm,
\qquad\text{and}\qquad
\bfn=\bfm,
\label{22}
\ee
whereby it follows that
\begin{align}
|\bfu_{\raisebox{-1.15pt}{$\scriptstyle\rho$}}\ctimes\bfxi_{\theta}+\bfxi_{\rho}\ctimes\bfu_{\raisebox{-1.15pt}{$\scriptstyle\theta$}}|^2
&-[\bfn\cdot(\bfu_{\raisebox{-1.15pt}{$\scriptstyle\rho$}}\ctimes\bfxi_{\theta}+\bfxi_{\rho}\ctimes\bfu_{\raisebox{-1.15pt}{$\scriptstyle\theta$}})]^2
\notag\\[4pt]
&\qquad=[\bfe\cdot(\bfu_{\raisebox{-1.15pt}{$\scriptstyle\rho$}}\ctimes\bfxi_{\theta}+\bfxi_{\rho}\ctimes\bfu_{\raisebox{-1.15pt}{$\scriptstyle\theta$}})]^2+[\bfeperp\cdot(\bfu_{\raisebox{-1.15pt}{$\scriptstyle\rho$}}\times\bfxi_{\theta}+\bfxi_{\rho}\ctimes\bfu_{\raisebox{-1.15pt}{$\scriptstyle\theta$}})]^2
\notag\\[4pt]
&\qquad=\rho^2(\bfm\cdot\bfu_{\rho})^2+(\bfm\cdot\bfu_{\theta})^2.
\label{24}
\end{align}
Also, bearing in mind the assumed smoothness of $\bfu$,
\begin{align}
{\bf0}&=\aintdim(\bfu_{\rho}\ctimes\bfu)_{\theta}\drhodtheta
\notag\\[4pt]
&=\aintdim(\bfu_{\rho\theta}\ctimes\bfu+\bfu_{\rho}\ctimes\bfu_{\theta})\drhodtheta
\notag\\[4pt]
&=\lint(\bfu_{\theta}\ctimes\bfu)|_{\rho\eql1}\dtheta
+\aintdim(\bfu_{\rho}\ctimes\bfu_{\theta}-\bfu_{\theta}\ctimes\bfu_{\rho})\drhodtheta
\notag\\[4pt]
&=\lint(\bfu_{\theta}\ctimes\bfu)|_{\rho\eql1}\dtheta+
2\aintdim\bfu_{\rho}\ctimes\bfu_{\theta}\drhodtheta,
\end{align}
the $\bfm$-component of which yields
\be
\aintdim\bfm\cdot(\bfu_{\rho}\ctimes\bfu_{\theta})\drhodtheta
=-\half\lint\bfm\cdot(\bfu_{\theta}\ctimes\bfu)|_{\rho\eql1}\dtheta.
\label{25}
\ee

On using \eqref{20}, \eqref{22}, \eqref{24}, and \eqref{25} in \eqref{18}, the second-variation condition at the trivial solution reduces to
\begin{multline}
\aintdim\nu\Big(\rho(\bfm \cdot \bfu_{\rho})^2 + \frac{1}{\rho}(\bfm \cdot \bfu_{\theta})^2 \Big)\drhodtheta
\\[4pt]
+\lint(\nu\mskip1mu\bfm \cdot (\bfu\ctimes\bfu_{\theta})+|\bfu_{\theta \theta}|^2-(1+\nu)|\bfu_{\theta}|^2)|_{\rho\eql1}\dtheta\ge0,
\label{26}
\end{multline}
while, at the trivial solution, the constraint condition \eqref{19} reads
\be
\bfeperp \cdot \bfu_{\theta}|_{\rho\eql1}=0.
\label{27}
\ee

To facilitate the derivation of pointwise conditions from \eqref{26}, introduce
\be
v = \bfe \cdot \bfu
\qquad\text{and}\qquad
w = \bfm \cdot \bfu,
\label{28}
\ee
which can be regarded, respectively, as radial and axial displacements from the trivial solution. Then, by \eqref{27} and the assumed smoothness of $\bfu$,
\be
\bfeperp\cdot\bfu_{\theta\theta}=\bfe \cdot \bfu_{\theta},\quad v_{\theta} = \bfeperp \cdot \bfu + \bfe \cdot \bfu_{\theta},\quad v_{\theta \theta} = \bfe \cdot(\bfu_{\theta \theta}-\bfu),
\label{29}
\ee
while, by \eqref{23} and \eqref{27},
\be
\lint\bfm \cdot (\bfu \ctimes \bfu_{\theta})\dtheta =  \lint(\bfe \ctimes \bfeperp) \cdot (\bfu \ctimes \bfu_{\theta})\dtheta = -\lint(\bfe \cdot \bfu_{\theta})(\bfeperp \cdot \bfu)\dtheta.
\label{30}
\ee
Next, in view of the assumed smoothness of $\bfu$, using \eqref{28} in the first term on the left-hand side of \eqref{26} yields
\be
\aintdim\nu\bigg(\rho(\bfm \cdot \bfu_{\rho})^2 + \frac{1}{\rho}(\bfm \cdot \bfu_{\theta})^2 \bigg)\drhodtheta
=\aintdim\nu\bigg( \rho w_{\rho}^2 + \frac{1}{\rho} w_{\theta}^2 \bigg)\drhodtheta,
\label{svid1}
\ee
while using \eqref{28}, \eqref{29}, and \eqref{30} in the second term on the left-hand side of \eqref{26} yields
\begin{multline}
\lint(\nu\mskip1mu\bfm \cdot (\bfu\ctimes\bfu_{\theta})+|\bfu_{\theta \theta}|^2-(1+\nu)|\bfu_{\theta}|^2)|_{\rho\eql1}\dtheta
\\[4pt]
=-\lint[\nu(\bfe \cdot \bfu_{\theta})(\bfeperp \cdot \bfu) - |\bfu_{\theta \theta}|^2 + (1+\nu) |\bfu_{\theta}|^2 ]_{\rho\eql1} \dtheta
\\[4pt]
=\lint [ w_{\theta \theta}^2 - (1+\nu) w_{\theta}^2 + (v_{\theta \theta} + v)^2 -\nu(v_{\theta}^2 -v^2)]_{\rho\eql1}\dtheta.
\label{svid2}
\end{multline}
With \eqref{svid1} and \eqref{svid2}, the inequality \eqref{26} expressing the second-variation condition at the trivial solution becomes
\begin{multline}
\aintdim\nu\bigg( \rho w_{\rho}^2 + \frac{1}{\rho} w_{\theta}^2\bigg)\drhodtheta
\\[4pt]
+  \lint [ w_{\theta \theta}^2 - (1+\nu) w_{\theta}^2 + (v_{\theta \theta} + v)^2 -\nu(v_{\theta}^2 -v^2)]_{\rho\eql1} \dtheta\ge 0.
\label{31}
\end{multline}

An immediate consequence of \eqref{31} is that the contributions of $v$ and $w$ to the second variation of the net free-energy are completely separated. Thus, \eqref{31} holds for all $\bfu$ if and only if the inequalities
\be
\aintdim\nu\bigg( \rho w_{\rho}^2 + \frac{1}{\rho} w_{\theta}^2 \bigg)\drhodtheta
+ \lint[w_{\theta \theta}^2 - (1+\nu) w_{\theta}^2]_{\rho\eql1} \dtheta \ge 0 \label{32} \ee
and
\be
\lint[(v_{\theta \theta} + v)^2 - \nu(v_{\theta}^2-v^2)]_{\rho\eql1} \dtheta \ge 0 \label{33}
\ee
hold independently.

Conditions necessary and sufficient to ensure satisfaction of the quadratic inequalities \eqref{32} and \eqref{33} can be determined via solving relevant variational problems. First, \eqref{32} is satisfied for all smooth $w$ if and only if it is satisfied for all $w$ consistent with the normalization condition
\be
\aintdim\rho w^2\drhodtheta= 1.
\label{normalization}
\ee
The left-hand side of \eqref{32} admits a minimum when subjected to the normalization condition \eqref{normalization}. Hence, \eqref{32} is satisfied if and only if the minimum is nonnegative. Any such minimum is determined by a solution of the Euler--Lagrange equations
\be
\left.
\begin{array}{c}
\displaystyle
\nu\bigg(w_{\rho \rho} +\frac{1}{\rho}w_{\rho} + \frac{1}{\rho^2} w_{\theta\theta}\bigg) + \eta w = 0, 
\cr\noalign{\vskip4pt}
[\nu w_{\rho} + w_{\theta\theta\theta\theta} + (1 + \nu)w_{\theta\theta}]_{\rho\eql1} = 0, 
\end{array}
\right\}
\label{ELeqns}
\ee
where $\eta$ is a Lagrange multiplier associated with \eqref{normalization}. At a solution $w$ of \eqref{ELeqns}, the value of the left-hand side of \eqref{32} is exactly $\eta$. Thus, \eqref{32} is satisfied if and only if $\eta \geq 0$ for every solution of \eqref{ELeqns}, or equivalently, \eqref{32} is violated if and only if $\eta < 0$ for at least one solution of \eqref{ELeqns}. Separable solutions of \eqref{ELeqns}$_1$ are provided by
\be
w(\rho,\theta) = P(\rho) (C_1 \cos n \theta + C_2 \sin n \theta), \label{solution}
\ee
where $n$ is a nonnegative integer and $P$ satisfies
\be P''(\rho) + \frac{1}{\rho} P'(\rho) + \bigg(\frac{\eta}{\nu} - \frac{n^2}{\rho^2}\bigg)P(\rho) = 0. \label{bessel}
\ee
By a change of variables, \eqref{bessel} becomes a Bessel equation of order $n$ for $\eta > 0$ or a modified Bessel equation of order $n$ for $\eta < 0$. For the second of these alternatives, a bounded solution of \eqref{bessel} is given by
\be P(\rho) = I_n\bigg(\sqrt{-\frac{\eta}{\nu}} \rho\bigg), \label{besselfunction}
\ee
where $I_n$ is the modified Bessel function of the first kind and order $n$, which admits a series representation of the form
\be
I_n(z) = \sum_{k=0}^{\infty} \frac{1}{k! (k + n)!}\bigg(\frac{z}{2}\bigg)^{\mskip-4mun+ 2k}.
\label{In}
\ee
Substituting \eqref{besselfunction} into \eqref{solution} and \eqref{ELeqns}$_2$ yields
\be
\nu x I_n'(x) + (n^4 - (1 + \nu) n^2) I_n(x) = 0, \label{120}
\ee
with $x = \sqrt{-\eta/\nu}$ being positive. It follows from \eqref{In} that
\be
x I_n'(x) > n I_n(x),
\ee
which, upon substituting into \eqref{120}, leads to
\be
n(n-1)(n^2+n-\nu)<0.
\ee
The last inequality holds only if $n \geq 2$ and $\nu > 6$. That is, a necessary condition for \eqref{bessel} to have a solution with negative $\eta$, and therefore, to violate \eqref{32}, is $\nu > 6$. In other words, a sufficient condition for \eqref{32}
to be satisfied is
\be
\nu\le6.
\label{34}
\ee
That this condition is also necessary for \eqref{32} to be satisfied becomes evident on using
\be
w(\rho,\theta)=\rho^2\sin(2\theta),
\label{35}
\ee
in \eqref{32}.

By a similar argument, it can be shown that inequality \eqref{33} holds if and only if $\eta \geq 0$ for every solution of
\be
v_{\theta \theta \theta \theta} + (2 + \nu) v_{\theta \theta} + (1 + \nu - \eta) v = 0.
\ee
This is the case if
\be\nu\le3, \label{36} \ee
which is thus sufficient condition for \eqref{33} to hold. This condition is also necessary, as becomes evident on choosing $v$ in \eqref{33} to be of the explicit form
\be
v(\rho,\theta)=V(\rho)\sin(2\theta),
\label{37}
\ee
where $V$ is differentiable and obeys
\be
V(0)=0
\ee
but is otherwise unrestricted.

Comparison of \eqref{34} and \eqref{36} leads to the conclusion that inequality \eqref{36} is a necessary condition for the trivial solution \eqref{20} to be stable. Notice that \eqref{37} describes a flat surface bounded by an ellipse-like loop.

\cite{gm} obtain $\nu=3$ ($\gamma=48\pi$) as the critical value at which a flat circular disc becomes unstable. This is achieved by examining the energy change induced by subjecting a flat circular disc to ``a small periodic displacement in the redial direction,'' which happens to coincide with \eqref{37}. In the present work, the stability condition \eqref{36} is derived by solving the second variation condition \eqref{31}, which eliminates the need for guesswork.

Notice that, in their (5.8), \cite{gm} obtain $\nu=6$ ($\gamma = 96 \pi$)  as ``the critical value of the bifurcation parameter for a supercritical pitchfork bifurcation.'' The argument leading to this conclusion involves evaluating the net free-energy \eqref{energy0} of a certain one-parameter family surfaces that represents ``the simplest parametrization of a twisted saddle.'' Computing and setting equal to zero the derivative of the resulting expression with respect to the underlying parameter ostensibly yields an equilibrium solution and, thus, information regarding a bifurcation condition. However, this claim is correct only if the family of surfaces actually includes an equilibrium solution. The extremum of a functional on a subset of the underlying function space may not generally correspond to an extremum of the functional over the entire space. \cite{gm} further note that this critical value is consistent with their ``exact linear stability analysis.'' However, their stability analysis, as described above, concludes that the disc configuration becomes unstable for $\gamma$ greater than $48\pi$, and hence must be unstable before reaching the critical value $96\pi$.

The bifurcation analysis presented in the next section will clarify some of these issues.

\section{Bifurcation from trivial solutions}
\label{bifurcation}

By the implicit function theorem \citep{golubitsky,chen}, the dimensionless boundary-value problem consisting of \eqref{12}, \eqref{16}, and \eqref{17} possesses a nontrivial solution branch bifurcating from the trivial solution branch \eqref{20} only if the linearized equations possesses a nontrivial solution. Let $\bfu$ and $\epsilon$ denote increments of $\bfxi$ and $\lambda$, respectively. On invoking the definition \eqref{14} of $\bfn$, a lengthy but straightforward calculation results in the  linearized version
\be
\bfm\cdot\bigg(\bfu_{\rho\rho}+\frac{1}{\rho}\bfu_\rho
+\frac{1}{\rho^2}\bfu_{\theta \theta}\bigg) = 0
\label{38}
\ee
of \eqref{16}, which is the Laplace equation of the $\bfm$-component of $\bfu$. Similarly, the linearized version of the boundary condition \eqref{17} has the form
\be
[(1+\nu)\bfu_{\theta\theta}
+\nu(\bfm\cdot\bfu_{\rho})\bfm
+\nu\bfu_{\theta}\ctimes\bfm
+\bfu_{\theta\theta\theta\theta}]_{\rho\eql1}
+\epsilon\bfe-\epsilon_{\theta}\bfeperp
=\bf0.
\label{39}
\ee
By the choice of notation, the increment $\bfu$ of $\bfxi$ also satisfies the constraint \eqref{27}.

Any solution $\bfu$ of \eqref{27}, \eqref{38}, and \eqref{39} admits a decomposition of the form
\be
\bfu(\rho,\theta)=w(\rho,\theta)\bfm+\bfv(\rho,\theta),
\label{40}
\ee
where $\bfv$ obeys
\be
\bfv\cdot\bfm=0
\label{perpendicularity}
\ee
and, thus, is confined to the plane spanned by $\bfe$ and $\bfeperp$, the notation being chosen in accord with \eqref{28}. Consistent with \eqref{40} and \eqref{perpendicularity}, $w$ and $\bfv$ are termed out-of-plane and in-plane displacements. The partial-differential equation \eqref{38} involves only the out-of-plane displacement. Also, the constraint \eqref{27} involves only the restriction of the in-plane displacement to the unit circle. It is therefore convenient to decompose the boundary condition \eqref{39} correspondingly and to write the boundary-value problem consisting of \eqref{27}, \eqref{38}, and \eqref{39} as two separate systems.

Of these systems, that governing the out-of-plane displacement $w$ consists of the scalar Laplace equation
\be
w_{\rho\rho}+\frac{1}{\rho}w_\rho+\frac{1}{\rho^2}w_{\theta\theta} = 0
\label{41}
\ee
and the linear boundary condition
\be
[(1+\nu)w_{\theta\theta}+\nu w_{\rho}+w_{\theta\theta\theta\theta}]_{\rho\eql1}= 0.
\label{42}
\ee
Like the general nonlinear boundary-value problem \eqref{12}, \eqref{16}, and \eqref{17} for $\bfxi$ and $\lambda$, the boundary-value problem \eqref{41}--\eqref{42} for $w$ involves a second-order partial differential equation subject to a boundary condition involving fourth-order derivatives, and in this sense is nonclassical.

The remaining system, for the restriction $\bfv|_{\rho\eql1}$ of vector-valued in-plane displacement $\bfv$ to the unit circle and the Lagrange multiplier increment $\epsilon$, consists of two linear ordinary-differential equations,
\be
\left.
\begin{array}{c}
[(1+\nu)\bfv_{\theta\theta}
-\nu(\bfe\cdot\bfv_{\theta})\bfeperp
+\bfv_{\theta\theta\theta\theta}]_{\rho\eql1}+\epsilon\bfe
-\epsilon_{\theta}\bfeperp
={\bf0},
\cr\noalign{\vskip4pt}
\bfeperp\cdot\bfv_{\theta}|_{\rho\eql1} = 0,
\end{array}
\right\}
\label{43}
\ee
which hold on the unit circle.

Consider first the out-of-plane problem. The Laplace equation \eqref{41} admits separable solutions
\be
w(\rho, \theta) =  P(\rho) \Theta(\theta).
\label{44}
\ee
Substituting \eqref{44} into \eqref{41}, solving the resulting ordinary-differential equations, and imposing the requirement that the out-of-plane displacement $w$ be smooth and bounded yields
\be
w(\rho,\theta) = C_1\rho^n\cos(n \theta) + C_2\rho^n\sin (n \theta),
\label{45}
\ee
where $C_1$ and $C_2$ are constants and $n$ is an arbitrary nonnegative integer. As defined in \eqref{45}, $w$ satisfies the boundary condition \eqref{42} if and only if
\be
n(n - 1)[\nu-n (n + 1)] = 0.
\label{46}
\ee
The root $n = 0$ corresponds to a rigid body translation in the direction of $\bfm$, and the root $n = 1$ corresponds to a rigid body rotation about a diameter of the unit disk. These solutions are of no physical interest. What remains are the possible bifurcation points
\be
\nu = n(n + 1),
\qquad
n\ge2.
\label{47}
\ee
In particular, the solution corresponding to the choice $n = 2$ is a hyperbolic paraboloid---a saddle-like surface. Although the out-of-plane bifurcation solutions determined by \eqref{45} and \eqref{47} are of mathematical interest, they are unstable and therefore not observable physically. Indeed, it is easily confirmed that the stability condition \eqref{36} is not satisfied at the bifurcation points \eqref{47}. The trivial solution is therefore unstable at these bifurcation points, as is any nontrivial solution that is sufficiently close to the unstable trivial solution. This is evident from the continuous dependence of the net free-energy \eqref{energy} on $\bfx$. Indeed, if the second variation, namely the left-hand side of \eqref{18dim}, is negative for a solution $(\bfx,\lambda)$ and a variation $\bfu$, it must remain negative for solutions in a small neighborhood of that solution.

In contrast to a claim of \cite{gm},  the above result shows that a transition from a flat circular disk to a stable saddle-like configuration is not possible, as such configurations are unstable according to the present analysis. The energy stability criterion asserts that a configuration is stable if the energy associated with it is lower than {\em all} neighboring configurations. Comparing the energy with those of a pre-selected set of configurations need not suffice to ensure stability.

Consider next the in-plane problem. The most general expressions for $\bfv|_{\rho\eql1}$ and $\epsilon$ consistent with the system \eqref{43} are
\begin{multline}
\bfv(1,\theta)=C_1 \bfeperp + C_2 (\bfe - \theta \bfeperp) + C_3[(\sin \theta) \bfe + (\cos \theta)\bfeperp] + C_4[(\cos \theta) \bfe - (\sin \theta) \bfeperp]
\\[4pt]+ C_5\big(\sqrt{1+\nu} \sin\big(\sqrt{1+\nu}\mskip2mu\theta\big) \bfe + \cos\big(\sqrt{1+\nu}\mskip2mu\theta\big) \bfeperp\big)
\\[4pt]+ C_6\big(\sqrt{1+\nu} \cos\big(\sqrt{1+\nu}\mskip2mu\theta\big) \bfe - \sin\big(\sqrt{1+\nu}\mskip2mu\theta\big) \bfeperp\big)
\label{48}
\end{multline}
and
\be
\epsilon(\theta) = C_2(2-\nu) - 3 C_5\nu\sqrt{1+\nu} \sin\big(\sqrt{1+\nu}\mskip2mu\theta\big) - 3 C_6\nu\sqrt{1+\nu} \cos\big(\sqrt{1+\nu}\mskip2mu\theta\big).
\label{49}
\ee
Of the terms in \eqref{48}, that with coefficient $C_1$ corresponds to a rigid-body rotation about the center of the disk, while those with coefficients $C_3$ and $C_4$ correspond to rigid-body translations in the plane of the disk. In view of the invariance \eqref{11}, rigid-body transformations are of no physical interest in the present context. Moreover, the closure condition $\bfv(1,0)=\bfv(1,2\pi)$ requires that $C_2=0$. Hence, the only remaining terms of nontrivial consequence are those with coefficients $C_5$ and $C_6$. A further requirement of the closure condition is that
\be \sqrt{1+\nu}=m, \ee
with $m$ being an arbitrary nonnegative integer. Since $\nu$ is positive, the possible bifurcation points are given by
\be
\nu=m^2-1,
\qquad m\ge2.
\label{51}
\ee
In view of the stability condition \eqref{36}, the only viable choice of $m$ in \eqref{51} is $m=2$. This yields $\nu=3$, which coincides the value of $\nu$ at which the trivial solution branch loses stability. It can therefore be argued that, as $\nu$ increases from some value less than 3, a flat disk becomes unstable at $\nu=3$ and bifurcates to a noncircular flat shape. With reference to the definition \eqref{nu} of $\nu$, this can be achieved in various ways. For instance, if $\sigma$ and $a$ are fixed, as would occur in a series experiments involving a particular batch of soap solution and loops made of a single type of fishing line starting with a length $L=2\pi R$ such that $\nu<3$ and gradually increasing that length to reach $\nu=3$.

\section{Summary}

In this work, the Euler--Plateau problem first studied by \cite{gm} is reformulated through a paremeterization $\bfx$ of the surface of the soap film and the bounding elastic loop. This formulation leads to a minimization problem for a net free-energy functional with some unusual features. Specifically, the functional involves the first-order derivatives of $\bfx$ in the interior of the surface but the second-order derivatives of $\bfx$ on the boundary.

By applying standard variational methods to this minimization problem, first and second variation conditions are derived in terms of $\bfx$. The first variation condition gives the equations of equilibrium, which form an ostensibly well-posed boundary-value-problem. The shapes of the soap film and the bounding loop can be determined by solving this boundary-value-problem. This represents a considerable advance, as previous work provided only an ad hoc derivation of the first variation condition involving the mean curvature of the spanning surface, the curvature and torsion of the bounding loop, and the angle between the principal normal of the Frenet frame of the bounding loop and the restriction to the bounding loop of the normal to the spanning surface. Despite their geometrically appealing nature, the corresponding equilibrium equations provide fewer equations than they involve unknowns. Moreover, using a solution to these equations to construct solutions to the Euler--Plateau problem would be hindered by the still unresolved closed curve problem discussed by \cite{e} and \cite{f}. The present analytical formulation of the Euler--Plateau problem circumvents these difficulties. It is shown that the equilibrium conditions obtained previously can be derived from the equations of equilibrium derived here. Further, rigorous connections between the equations of equilibrium and the force and moment balances are established.

The second variation condition, which is not previously derived, gives a stability condition in the form of an integral inequality. By solving this integral inequality, pointwise stability conditions are derived for the flat circular solution, with respect to noncircular in-plane perturbations and to out-of-plane perturbations.

A rigorous bifurcation analysis is provided. That analysis identifies the bifurcation points from the flat circular solution branch to flat noncircular solution branches, and to nonplanar solution branches. Combining with the stability analysis, this gives a comprehensive understanding of the evolution of the shape of spanning surface as the bifurcation parameter varies. It also provides rigorous justification of some correct results previously obtained through ad hoc arguments, and points out the previous results that are incorrect.

Possible extensions of the Euler--Plateau problem include (i) a more general form of the free energy for the spanning surface that includes, in addition to surface tension, elastic stretches and bending effects; (ii) a more general form of the energy for the bounding loop that includes intrinsic curvature and/or intrinsic twist; (iii) a comprehensive stability and bifurcation analysis for flat noncircular solutions and for nonplanar solutions. Some of these tasks are in progress.

\bibliographystyle{elsarticle-harv}

\end{document}